\documentclass[aps,pra,twocolumn,amssymb,amsfonts,floatfix,superscriptaddress,longbibliography,notitlepage,nofootinbib]{revtex4-1}
\usepackage[utf8]{inputenc}
\usepackage[draft,inline,nomargin,index,marginclue]{fixme} 
\pdfoutput=1
\usepackage{color}
\usepackage{float}
\usepackage{xcolor}
\usepackage{mathtools,amsmath,amsthm,amsfonts,amssymb}
\usepackage{commath}
\usepackage{physics}
\usepackage{bm}
\usepackage[pdftex]{graphicx}
\usepackage[colorlinks=false]{hyperref}
\usepackage{enumitem}
\begin{document}

\title{Analysis of Chaos and Regularity in the Open Dicke Model}

\author{David Villase\~nor}
\author{Pablo Barberis-Blostein}
\affiliation{Instituto de Investigaciones en Matem\'aticas Aplicadas y en Sistemas, Universidad Nacional Aut\'onoma de M\'exico, C.P. 04510 CDMX, Mexico}


\begin{abstract}
We present an analysis of chaos and regularity in the open Dicke model, when dissipation is due to cavity losses. Due to the infinite Liouville space of this model, we also introduce a criterion to numerically find a complex spectrum which approximately represents the system spectrum. The isolated Dicke model has a well-defined classical limit with two degrees of freedom. We select two case studies where the classical isolated system shows regularity and where chaos appears. To characterize the open system as regular or chaotic, we study regions of the complex spectrum taking windows over the absolute value of its eigenvalues. Our results for this infinite-dimensional system agree with the Grobe-Haake-Sommers (GHS) conjecture for Markovian dissipative open quantum systems, finding the expected 2D Poisson distribution for regular regimes, and the distribution of the Ginibre unitary ensemble (GinUE) for the chaotic ones, respectively.
\end{abstract}

\maketitle

\section{INTRODUCTION}
\label{sec:Introduction}

The way to characterize the chaotic behavior in isolated quantum systems comes from the classical realm. Classically, the concept of chaos is explained as a strong sensibility to initial conditions. This sensibility is typically measured with the Lyapunov exponent, a rate of divergence between two initial trajectories which separate between them as time evolves~\cite{OttBook}. The extension of the last idea cannot be directly made in the quantum realm, due to the nature of quantum mechanics. Instead, the spectral fluctuations of the quantum system Hamiltonian have traditionally been studied through statistical tests of their eigenvalue spacings~\cite{HaakeBook,Guhr1998}.

For integrable (regular) quantum systems, the eigenvalue spacings generically follow the Poisson distribution associated with uncorrelated levels, as stated by the Berry-Tabor conjecture~\cite{Berry1977}. However, for nonintegrable (chaotic) quantum systems with time-reversal symmetry, the eigenvalue spacings follow the Wigner-Dyson distribution (Wigner-Dyson surmise) associated with level repulsion. The last was conjectured by Bohigas, Giannoni, and Schmit for quantum systems whose classical limit is chaotic~\cite{Bohigas1984}, and whose spectral fluctuations are described by the Gaussian orthogonal ensemble (GOE) of the random matrix theory~\cite{HaakeBook,MehtaBook,Guhr1998}. The last characterization is also applicable for systems without a well-defined classical limit~\cite{Hsu1993,Zelevinsky1996PR,Santos2010}.

When dissipation is taken into consideration, the system of interest can be modeled by an effective non-Hermitian Hamiltonian. Thus, chaotic behavior can be studied using its complex spectrum. This approach is well-known and widely used in mesoscopic and nuclear physics~\cite{Sokolov1988,Sokolov1989,Sokolov1992,Mizutori1993,Celardo2007PRE,Celardo2007PLB,Sorathia2009,Auerbach2011}. Another possibility to study dissipation is under the formalism of open quantum systems~\cite{BreuerBook,CarmichaelBook1993,CarmichaelBook2002,Manzano2020}, using a Lindblad master equation in the Markov approximation.

In the formalism of open quantum systems, the system state is given by an operator, the density operator, which acts on the Hilbert space of the system. The dynamics of the system state is dictated by a new operator, called Liouvillian superoperator or simply Liouvillian, which acts on the space of operators or Liouville space. The Liouvillian is in general a non-Hermitian operator with complex eigenvalues~\cite{BreuerBook,Ashida2020}. The spectrum of the Liouvillian does not have the same interpretation as the real spectrum of Hermitian Hamiltonians. This fact prevents a simple generalization of the criteria that characterize chaotic behavior in isolated quantum systems to open quantum systems. In this work, we study a quantum optical system, the open Dicke model, where atoms and photons interact in presence of dissipation. Thereby, due to its broad application in quantum optics, we follow the formalism of open quantum systems, using a Lindblad master equation in the Markov approximation, to characterize chaos in this system.

Pioneering studies trying to understand the chaotic nature of open quantum systems were performed in periodically kicked dissipative tops with classical limit~\cite{Grobe1988}, where it was found that the distribution of the complex-eigenvalue spacings, understood as the Euclidean distance in the complex plane, follows a 2D (two-dimensional) Poisson distribution when the classical model is regular. The last point is understandable at some extent, since the intuitive extrapolation from isolated regular systems suggests that the eigenvalue spacings must be uncorrelated in the plane. In contrast, when the classical model is chaotic, the eigenvalue spacing distribution was found to agree with the distribution of the Ginibre unitary ensemble (GinUE)~\cite{Ginibre1965}, showing a cubic level repulsion~\cite{Grobe1988,Grobe1989}.

The extrapolation of the results found in the periodically kicked dissipative tops to any open quantum system is nowadays called the Grobe-Haake-Sommers (GHS) conjecture~\cite{Grobe1988,Grobe1989}. It has been shown to be satisfied in other open quantum systems with finite dimension, as spin chains~\cite{Akemann2019,Hamazaki2020}, or Richardson-Gaudin Liouvillians~\cite{Rubio2022}, and seems to be universal. Some quantum systems in the chaotic case, using the effective non-Hermitian Hamiltonian approach to treat dissipation, show an eigenvalue spacing distribution that agrees with the GinUE distribution from the GHS conjecture~\cite{Jaiswal2019,Hamazaki2019}. Nevertheless, for nuclear and mesoscopic systems, the eigenvalue spacing distribution can be obtained analytically and deviates from the GinUE distribution~\cite{Sokolov1988,Sokolov1989,Sokolov1992,Mizutori1993}, suggesting that the GHS conjecture is not well established for all systems under this approach.

In this way, as the GHS conjecture has been validated for finite-dimensional systems following the formalism of open quantum systems, a natural question is to ask if it remains valid for infinite-dimensional systems. Thus, the main goal in this work is to study the open Dicke model, which is an infinite-dimensional system, and verify that it satisfies the GHS conjecture of spectral universality.

The isolated Dicke model represents the simplest interacting radiation-matter system~\cite{Dicke1954,Garraway2011,Kirton2019}. It was introduced to study the radiation process from a quantum mechanical perspective~\cite{Dicke1954,Hepp1973a,Hepp1973b,Wang1973,Gross1982,Garraway2011,Kirton2019}. In recent years it has been used in a broad variety of theoretical studies, including quantum phase transitions~\cite{Emary2003,Emary2003PRL,Romera2012,Brandes2013}, classical and quantum chaos~\cite{Deaguiar1992,Bastarrachea2014b,Bastarrachea2015,Bastarrachea2016PRE,Bastarrachea2017,Chavez2016,Lobez2016,Sinha2020}, quantum scarring~\cite{Deaguiar1991,Furuya1992,Bakemeier2013,Sinha2020,Pilatowsky2021NatCommun,Pilatowsky2021}, quantum localization in phase space~\cite{Wang2020,Villasenor2021,Pilatowsky2022}, nonequilibrium quantum dynamics~\cite{Altland2012NJP,Kloc2018,Lerma2018,Lerma2019,Kirton2019,Villasenor2020}, evolution of out-of-time-ordered correlators (OTOCs)~\cite{Chavez2019,Lewis-Swan2019,Pilatowsky2020,Kirkova2023}, connections between chaos, entanglement~\cite{Furuya1998,Lobez2016,Sinha2020}, and thermalization~\cite{Ray2016,Kirkova2023,Villasenor2023}, among others.

The Dicke model can be experimentally realized with setups as diverse as superconducting circuits~\cite{Jaako2016}, cavity assisted Raman transitions~\cite{Baden2014,Zhang2018}, trapped ions~\cite{Cohn2018,Safavi2018}, and others. Moreover, this model has a well-defined classical limit with two degrees of freedom~\cite{Deaguiar1992,Chavez2016}, which depending on the parameters and energy regions, can show regular or chaotic motion.

A Dicke model including cavity dissipation or collective atomic dissipation is known as the open Dicke model. This model has been studied from superradiance and quantum phase transitions~\cite{Dimer2007,Kirton2018,Gelhausen2018,Carollo2021,Boneberg2022,Jaeger2022}, to classical and quantum chaos~\cite{Stitely2020,Prasad2022,Ray2022}. Some studies have focused in a particular version of the model, as the two-photon open Dicke model~\cite{Garbe2020,Li2022}. Experimental realizations of the open Dicke model with optical cavities are shown in Refs.~\cite{Zhiqiang2017,Klinder2015PNAS}. Moreover, the study of this model could be extended to more general dissipation channels, as those including collective atomic decay or considering temperature effects~\cite{Kirton2019,Roses2020,BreuerBook,CarmichaelBook1993,CarmichaelBook2002}.

The open Dicke model has an infinite Liouville space, this makes the study of its spectrum difficult, since we need to find a numerically approximated complex spectrum that represents the original one. In this regard, in this work we first propose a criterion for finding meaningful eigenstates and eigenvalues of the system. Then, we apply the standard methodology of open quantum systems to reveal the appearance of chaotic behavior in the open Dicke model using the GHS conjecture.

The article is organized as follows. In Sec.~\ref{sec:OpenDickeModel}, we introduce the isolated Dicke model and discuss its important features. Next, we introduce the open Dicke model using a Lindblad master equation in the Markov approximation and discuss relevant results as the dissipative phase transition. In Sec.~\ref{sec:ConvergenceOpenDickeModel}, we propose a convergence criterion for the eigenstates and eigenvalues of the Dicke Liouvillian. In Sec.~\ref{sec:SpectralAnalysis}, we outline the standard procedures for performing spectral analysis in open quantum systems, along with a brief review of spectral analysis in isolated quantum systems. The main results of the article concerning chaos and regularity in the open Dicke model are shown in Sec.~\ref{sec:ChaosRegularityOpenDickeModel}. Finally, we summarize our results and present our conclusions in Sec.~\ref{sec:Conclusions}.

\section{OPEN DICKE MODEL}
\label{sec:OpenDickeModel}

The Dicke model, which describes the interaction between a set of $\mathcal{N}$ two-level atoms and a single-mode electromagnetic field without dissipation (isolated system), is modeled with the Hamiltonian (setting $\hbar=1$)~\cite{Dicke1954}
\begin{equation}
    \hat{H}_{\text{D}} = \omega\hat{a}^{\dagger}\hat{a}+\omega_{0}\hat{J}_{z}+\frac{\gamma}{\sqrt{\mathcal{N}}}(\hat{a}^{\dagger}+\hat{a})(\hat{J}_{+}+\hat{J}_{-}),
\end{equation}
where $\hat{a}^{\dagger}$ ($\hat{a}$) is the bosonic creation (annihilation) operator of the field mode. The set of operators $\{\hat{a}^{\dagger},\hat{a},\hat{\mathbb{I}}\}$ satisfy the (Heisenberg-Weyl) HW(1) algebra. Moreover, $\hat{J}_{+}$ ($\hat{J}_{-}$) is the raising (lowering) collective pseudospin operator, defined as $\hat{J}_{\pm}=\hat{J}_{x}\pm i\hat{J}_{y}$, where $\hat{J}_{\mu}=(1/2)\sum_{k=1}^{\mathcal{N}}\hat{\sigma}_{\mu}^{(k)}$ ($\mu=x,y,z$) are the collective pseudospin operators and $\hat{\sigma}_{\mu}$ are the Pauli matrices. The set of operators $\{\hat{J}_{+},\hat{J}_{-},\hat{J}_{z}\}$ satisfy the $SU(2)$ algebra in the same way as the Pauli matrices.

The Dicke Hamiltonian can be studied in invariant subspaces specified by the eigenvalues $j(j+1)$ of the squared total pseudospin operator $\hat{\textbf{J}}^{2}=\hat{J}_{x}^{2}+\hat{J}_{y}^{2}+\hat{J}_{z}^{2}$. We use the totally symmetric subspace, which is defined by the maximum pseudospin value $j=\mathcal{N}/2$ and includes the ground state. Furthermore, the Dicke Hamiltonian possesses a parity symmetry, $[\hat{H}_{\text{D}},\hat{\Pi}]=0$, where $\hat{\Pi} = \text{exp}[i\pi(\hat{a}^{\dagger}\hat{a}+\hat{J}_{z}+j\hat{\mathbb{I}})]$ is the parity operator. When an eigenbasis of this operator is selected as diagonalization basis of the Hamiltonian, the states of the system can be identified in two sectors of well-defined parity.

The main parameters of the Dicke Hamiltonian are the radiation frequency of the single-mode electromagnetic field, $\omega$, the atomic transition frequency from the ground state to the first excited state, $\omega_{0}$, and the coupling strength, $\gamma$, which modules the atom-field interaction within the system and reaches the critical value $\gamma_{\text{c}}=\sqrt{\omega\omega_{0}}/2$. At this value, the system develops a quantum phase transition going from a normal ($\gamma<\gamma_{\text{c}}$) to a superradiant ($\gamma>\gamma_{\text{c}}$) phase~\cite{Hepp1973a,Hepp1973b,Wang1973,Emary2003}. Classically, the Dicke model displays regular or chaotic behavior depending on the latter Hamiltonian parameters $(\omega,\omega_{0},\gamma)$ and excitation energies~\cite{Chavez2016}.

To study the open version of the Dicke model we use the Lindblad master equation in the Markov approximation. This equation describes the evolution of a small system of interest interacting with its environment~\cite{BreuerBook,CarmichaelBook1993,CarmichaelBook2002}. The Lindblad equation is obtained by assuming that the total system state is approximately always separable (Born approximation) and that its evolution does not depend on its past (Markov approximation). These approximations are justified when the system-environment coupling strength is weak.

When dissipation in the system is given by cavity losses, the open Dicke model is modeled with the following Lindblad master equation (setting $\hbar=1$)~\cite{BreuerBook,CarmichaelBook1993,CarmichaelBook2002}
\begin{equation}
    \frac{d \hat{\rho}}{d t} = \hat{\mathcal{L}}_{\text{D}}\hat{\rho} = -i[\hat{H}_{\text{D}},\hat{\rho}] + \kappa(2\hat{a}\hat{\rho}\hat{a}^{\dagger}-\{\hat{a}^{\dagger}\hat{a},\hat{\rho}\}),
\end{equation}
where $\hat{\rho}$ is the density matrix operator of the system, $\hat{\mathcal{L}}_{\text{D}}$ is the Liouvillian superoperator or Dicke Liouvillian, which acts over states in the Liouville space (operators acting on the Hilbert space of the system), $\kappa$ is the cavity decay coupling, and $\hat{a}^{\dag}$ ($\hat{a}$) is the bosonic creation (annihilation) operator of the field mode. The Dicke Liouvillian inherits a weak-parity symmetry of the Hamiltonian~\cite{Buca2012,Albert2014,Lieu2020}, since $[\hat{\mathcal{L}}_{\text{D}},\hat{\mathcal{P}}]=0$, where $\hat{\mathcal{P}}\hat{\rho}=\hat{\Pi}\hat{\rho}\hat{\Pi}^{\dagger}$ is the parity superoperator. An eigenbasis of this superoperator, used as diagonalization basis of the Liouvillian, identifies states of the system with well-defined parity analogously to the isolated system.

When cavity dissipation is considered in the open Dicke model, a quantum dissipative phase transition takes place at the critical coupling strength~\cite{Dimer2007,Kirton2019,Roses2020}
\begin{equation}
    \gamma_{\text{c}}^{\text{os}}=\frac{\sqrt{\omega\omega_{0}}}{2}\sqrt{1+\frac{\kappa^{2}}{\omega^{2}}},
\end{equation}
defining, analogously to the isolated system, two phases in the open system, a normal ($\gamma<\gamma_{\text{c}}^{\text{os}}$) and a superradiant ($\gamma>\gamma_{\text{c}}^{\text{os}}$) dissipative phase, respectively.

In this work, we use dimensionless Hamiltonian parameters scaled to the cavity decay coupling $\kappa$, $(\widetilde{\omega},\widetilde{\omega_{0}},\widetilde{\gamma})=(\omega/\kappa,\omega_{0}/\kappa,\gamma/\kappa)$. For convenience, from now on we remove the tilde in the last scaled parameters. We choose the resonant frequency case $\omega=\omega_0=1$, such that, the critical coupling strength value of the isolated and open system is $\gamma_{\text{c}}=0.5$ and $\gamma_{\text{c}}^{\text{os}}=1/\sqrt{2}\approx0.707$, respectively. With the selected parameters, we consider two case studies, one with a coupling strength in the normal phase ($\gamma=0.2$), and another one in the superradiant phase ($\gamma=1$) of the open system. For these values, the classical isolated system shows regular and chaotic motion, respectively~\cite{Chavez2016}. Moreover, to perform the spectral analysis we select the sector of eigenstates (eigenvalues) with positive parity in the Liouville space (see Appendix~\ref{app:DiagonalizationDickeLiouvillian} for an explanation of the Dicke Liouvillian with well-defined parity). We choose the smallest system size $j=1$ ($\mathcal{N}=2$ atoms) to show the convergence criterion for eigenstates and eigenvalues of the Dicke Liouvillian and to later perform the spectral analysis related with chaos.

\section{CONVERGENCE OF EIGENVALUES AND EIGENSTATES OF THE OPEN DICKE MODEL}
\label{sec:ConvergenceOpenDickeModel}

The isolated Dicke model has an infinite-dimensional Hilbert space composed by a finite atomic subspace with dimension $2j+1$, and an infinite-dimensional bosonic subspace. To numerically find the eigenvalues and eigenstates of the Dicke Hamiltonian, a truncation of the Fock basis for the bosonic subspace by a finite value of the number of photons $n_{\max}$ is done, generating in this way a finite Hilbert space with dimension $\mathcal{D}_{\text{H}}=(2j+1)(n_{\max}+1)$. The space of operators acting on the Hilbert space, known as the Liouville space, is also of
infinite dimension for the Dicke model. A finite Liouville space can be obtained using the truncated basis of the Hilbert space, where the Liouville basis is composed by all the projectors of the aforementioned basis. Thus, the Liouville space dimension is the square of the Hilbert space dimension $\mathcal{D}_{\text{L}}=\mathcal{D}_{\text{H}}^{2}$. In this section we introduce a convergence criterion to find eigenstates and its corresponding eigenvalues that are numerically close to the eigenstates of operators acting on the Liouville space.

\subsection{Convergence of Eigenvalues}
\label{sec:ConvergenceEigenvalues}

A usual way to define convergence of eigenvalues in infinite-dimensional spaces is comparing the change of the eigenvalues $\epsilon_{k}$ for two truncation values, $n_{\max}$ and $n_{\max}+1$,
\begin{equation}
    \label{eqn:EigenvalueConvergence}
    \Delta\epsilon_{k}=|\epsilon_{k}^{n_{\max}+1}-\epsilon_{k}^{n_{\max}}| \leq \varepsilon,
\end{equation}
where $\varepsilon$ is a tolerance value. Thus, the eigenvalue $\epsilon_{k}$ is rejected when the change $\Delta\epsilon_{k}$ exceeds the threshold $\varepsilon$.

This method has been successfully tested in the eigenvalues of the Dicke Hamiltonian~\cite{Bastarrachea2014PSa}, but the implementation becomes computing demanding when the truncation size of the Hamiltonian matrix increases, since two diagonalizations are needed. For this reason, an alternative convergence criterion based on the eigenstates of the truncated Hamiltonian matrix was proposed, showing an equivalence with the eigenvalue convergence criterion and using only one diagonalization of the system~\cite{Bastarrachea2014PSb}.

When extending the eigenvalue convergence criterion to the Dicke
Liouvillian, we found it is not applicable. It is well known that for
complex non-Hermitian matrices, there is always a set of missing eigenvalues  in the complex plane~\cite{Kerner1985}. This result implies that there is no natural way to identify corresponding eigenvalues between matrices of different dimensions and Eq.~(\ref{eqn:EigenvalueConvergence}) can not be used.

\subsection{Convergence of Eigenstates}
\label{sec:ConvergenceEigenstates}

Since the eigenvalue convergence criterion [see Eq.~\eqref{eqn:EigenvalueConvergence}] is not applicable to eigenvalues of the Dicke Liouvillian, we propose in this work an extension of the eigenstate convergence criterion from the isolated Dicke model for its open version. A detailed description of the convergence criterion for eigenstates of the Dicke Hamiltonian and supporting studies are presented in Ref.~\cite{Bastarrachea2014PSb}. See Appendix~\ref{app:ConvergenceCriterionEigenstates} for explicit details about the extension of this convergence criterion to eigenstates of the Dicke Liouvillian. Next, we describe the overall idea, which consists in expanding the eigenstates of the Dicke Liouvillian $|\lambda_{k}\rangle\rangle$, which satisfy the eigenvalue equation
\begin{equation}
    \hat{\mathcal{L}}_{\text{D}}|\lambda_{k}\rangle\rangle=\lambda_{k}|\lambda_{k}\rangle\rangle,
\end{equation}
in a diagonalization basis, say the Liouville basis $|n',m'_{z};n,m_{z}\rangle\rangle=|n';j,m'_{z}\rangle\langle n;j,m_{z}|$ with $m'_{z},m_{z}=-j,-j+1,\ldots,j-1,j$ and $n',n=0,1,\ldots,n_{\max}$, where $n_{\max}$ identifies a truncation value of the bosonic subspace or maximum number of photons [see Eqs.~\eqref{eqn:LiouvilleBasis} and~\eqref{eqn:LiouvilleEigenstateExpansion}]. Now, it is possible to obtain two weight distributions by projecting the eigenstate wave function over the Liouville basis [see Eqs.~\eqref{eqn:EigenstateWaveFunction1} and~\eqref{eqn:EigenstateWaveFunction2}]. When the last distributions are evaluated at the truncation value $n'=n=n_{\max}$, it is expected that the contribution to the eigenstate wave function will be negligible
\begin{align}
    \label{eqn:ConvergenceCriterion1}
    P_{1,n_{\max}}^{k} = \sum_{n=0}^{n_{\max}}\sum_{m'_{z},m_{z} = -j}^{j}|c_{n_{\max},m'_{z},n,m_{z}}^{k}|^{2} \leq \Delta, \\
    \label{eqn:ConvergenceCriterion2}
    P_{2,n_{\max}}^{k} = \sum_{n'=0}^{n_{\max}}\sum_{m'_{z},m_{z} = -j}^{j}|c_{n',m'_{z},n_{\max},m_{z}}^{k}|^{2} \leq \Delta,
\end{align}
where
$c_{n',m'_{z},n,m_{z}}^{k}=\langle\langle
n',m'_{z};n,m_{z}|\lambda_{k}\rangle\rangle$ are the eigenstate wave function components and $\Delta$ is a tolerance value. The last equations mean that we can expand, with a degree of approximation given by $\Delta$, the eigenstate wave function of the complete Liouville space using the basis of the truncated space.

\begin{figure}[ht]
\centering
\includegraphics[width=\columnwidth]{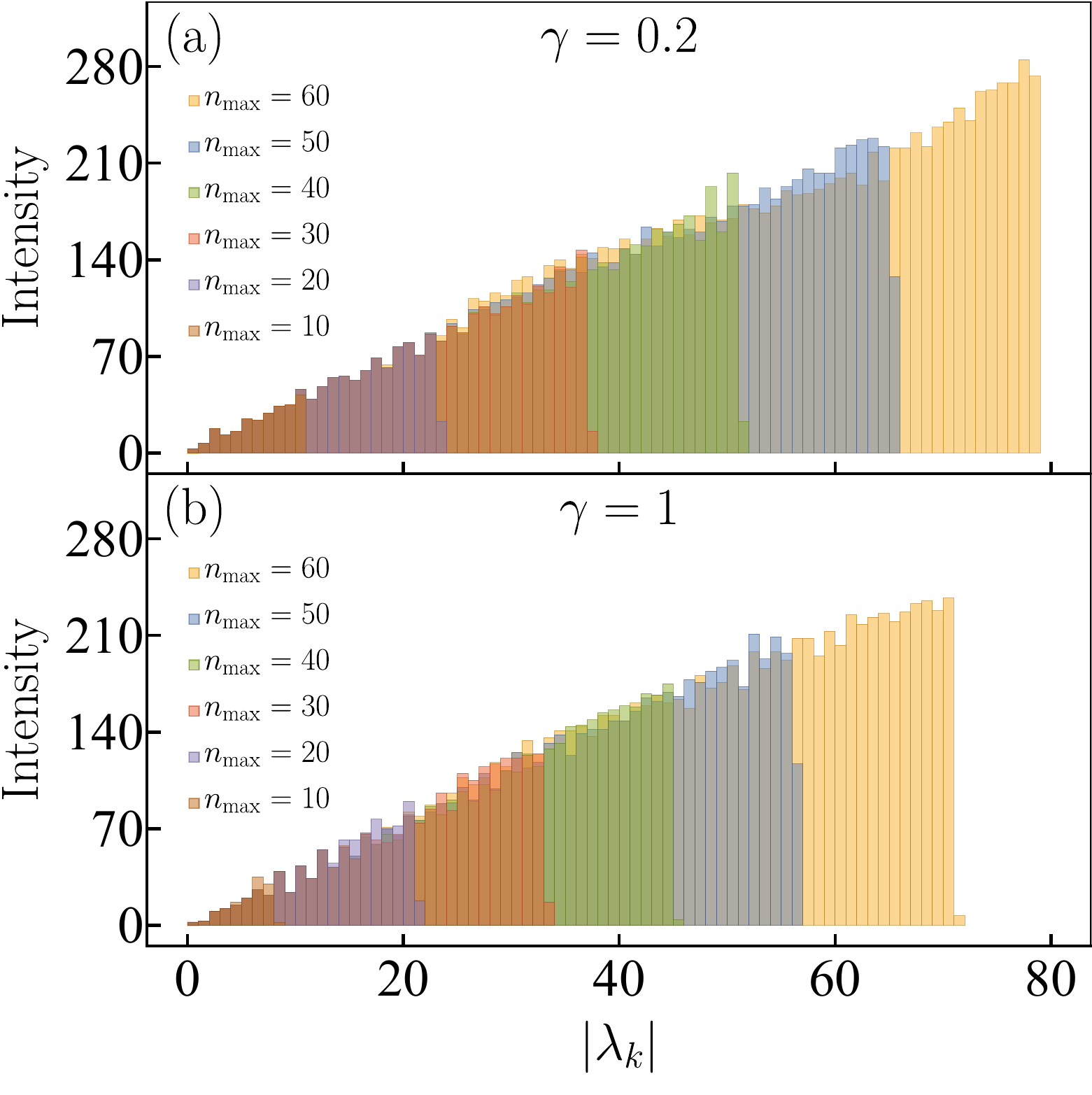}
\caption{(a) Histogram of the absolute value of the converged eigenvalues $|\lambda_{k}|$ with positive parity of the Dicke Liouvillian for the coupling strength $\gamma=0.2$, which were computed numerically for the truncation values $n_{\max}=10,20,30,40,50,60$. (b) The same as panel (a) for the coupling strength $\gamma=1$. The Liouvillian parameters in both panels are $\omega=\omega_{0}=j=\kappa=1$ and the tolerance value was chosen as $\Delta=10^{-3}$.}
\label{fig:ConvergedEigenvalues}
\end{figure}

\subsection{Convergence of Eigenstates vs. Eigenvalues}
\label{sec:ConvergenceEigenstatesEigenvalues}

In this section, we show numerical approximations for the eigenstates of the Dicke Liouvillian using the convergence criterion described in the previous section. We also argue why their corresponding eigenvalues are a good approximation to the eigenvalues of the full operator.

We select a set of truncation values $n_{\max}=10,20,30,40,50,60$ and diagonalize the Dicke Liouvillian for the parameters $\omega=\omega_0=j=\kappa=1$. We perform this procedure for two cases $\gamma=0.2$ and $\gamma=1$ to ensure we are diagonalizing the system in the normal ($\gamma<\gamma_{\text{c}}^{\text{os}}$) and superradiant ($\gamma>\gamma_{\text{c}}^{\text{os}}$) dissipative phase, respectively. We use the Liouville basis with positive parity ($P=+1$) to perform the convergence analysis. See Appendix~\ref{app:DiagonalizationDickeLiouvillian} for a complete description on how to diagonalize the Dicke Liouvillian using the Liouville basis and how to select the basis with well-defined parity. The well-converged sets of eigenvalues obtained with the convergence criterion will be used in Sec.~\ref{sec:ChaosRegularityOpenDickeModel} to perform the spectral analysis.

Now, we consider that the eigenvalue $\lambda_{k}$ is close to the eigenvalue of the full operator, when its corresponding eigenstate $|\lambda_{k}\rangle\rangle$ fulfills relations~\eqref{eqn:ConvergenceCriterion1} and~\eqref{eqn:ConvergenceCriterion2} simultaneously. In this case we say that $\lambda_{k}$ is well converged. We now take the converged eigenvalues for the set $n_{\max}=10,20,30,40,50,60$, and present the histogram of their absolute value $|\lambda_{k}|$ in Fig.~\ref{fig:ConvergedEigenvalues} for both cases $\gamma=0.2,1$. These histograms can be interpreted as a density of states for the absolute value $|\lambda_{k}|$. In both cases $\gamma=0.2,1$, we find that increasing the truncation value $n_{\max}$ preserves the behavior of the density of states. Furthermore, we see a number of converged eigenvalues higher for the low coupling strength $\gamma=0.2$ than for the high one $\gamma=1$. The last finding is intuitive extrapolating from the isolated system. In general, there are less converged eigenstates (eigenvalues) when the coupling strength is high in the system, since the eigenstate wave functions are more spread in the diagonalization basis and the convergence criterion is more difficult to be fulfilled~\cite{Bastarrachea2014PSa,Bastarrachea2014PSb}.

\begin{figure*}[ht]
\centering
\includegraphics[width=\textwidth]{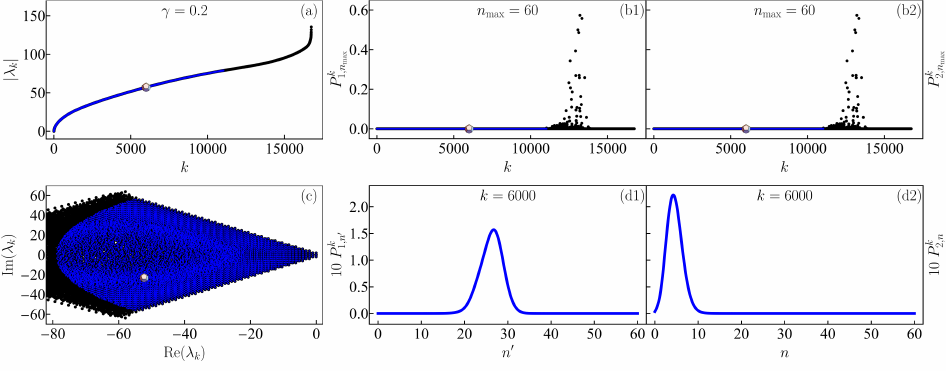}
\caption{(a) Absolute value of the eigenvalues $|\lambda_{k}|$ (black dots) with positive parity  of the Dicke Liouvillian for the coupling strength $\gamma=0.2$, which were computed numerically for the truncation value $n_{\max}=60$. The blue dots represent the well-converged eigenvalues selected under the eigenstate convergence criterion, and the diamond selects the eigenvalue with label $k=6000$, $\lambda_{6000}$. (b1, b2) Convergence criterion for the eigenstates $|\lambda_{k}\rangle\rangle$ with positive parity  of the Dicke Liouvillian [see Eqs.~\eqref{eqn:ConvergenceCriterion1}~and~\eqref{eqn:ConvergenceCriterion2}]. (c) Complex spectrum of the Dicke Liouvillian. (d1, d2): Wave function projections of the selected eigenstate $|\lambda_{6000}\rangle\rangle$ [see Eqs.~\eqref{eqn:EigenstateWaveFunction1}~and~\eqref{eqn:EigenstateWaveFunction2}]. The Liouvillian parameters in all panels are $\omega=\omega_{0}=j=\kappa=1$ and the tolerance value was chosen as $\Delta=10^{-3}$.}
\label{fig:RegularSpectrum}
\end{figure*}

\begin{figure*}[ht]
\centering
\includegraphics[width=\textwidth]{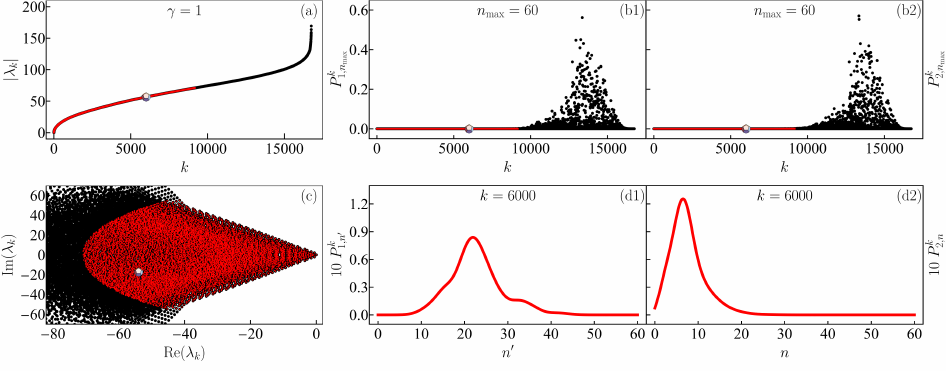}
\caption{The same as Fig.~\ref{fig:RegularSpectrum} for the coupling strength $\gamma=1$.}
\label{fig:ChaoticSpectrum}
\end{figure*}

To visualize explicitly the convergence criterion in the eigenstate wave functions, we focus on the single truncation value $n_{\max}=60$, for which we obtained $N_{\text{CES}}=11030,9165$ converged eigenstates (eigenvalues) for $\gamma=0.2,1$ with a tolerance value $\Delta=10^{-3}$. In Fig.~\ref{fig:RegularSpectrum} we show the case $\gamma=0.2$, where the complex spectrum ordered by the eigenvalue absolute value $|\lambda_{k}|$ is presented in Fig.~\ref{fig:RegularSpectrum}~(a). In this panel, the black dots represent the complete set of eigenvalues, while the blue dots represent the converged ones.

Figures~\ref{fig:RegularSpectrum}~(b1) and~\ref{fig:RegularSpectrum}~(b2) show the convergence criterion computed for all eigenstates $|\lambda_{k}\rangle\rangle$ [see Eqs.~\eqref{eqn:ConvergenceCriterion1}~and~\eqref{eqn:ConvergenceCriterion2}]. In the same way, the black dots represent the criterion computed for
all the set of eigenstates, while the blue dots for the converged ones. In these panels, a fraction of eigenstates for which the criterion is apparently fulfilled can be seen beyond $k=15000$ and even for the region $k>10000$, where the lack of convergence arises. Nevertheless, to avoid ambiguities by selecting them, and recalling that they are ordered by the increasing eigenvalue absolute value $|\lambda_{k}|$, we select them until the first eigenstate does not fulfill the criterion, discarding the remaining ones.

In Fig.~\ref{fig:RegularSpectrum}~(c) we show the spectrum in the complex plane. In this panel and panels (a) and (b1)-(b2), a 3D diamond, representing a particular eigenvalue (eigenstate) with label $k=6000$, is shown. In Figs.~\ref{fig:RegularSpectrum}~(d1) and~\ref{fig:RegularSpectrum}~(d2) we show the wave function of this selected eigenstate projected over the Liouville basis [see Eqs.~\eqref{eqn:EigenstateWaveFunction1}~and~\eqref{eqn:EigenstateWaveFunction2}], where we can see that the wave function is completely contained in the truncated Liouville space for both projections.

We repeat the last analysis for the case $\gamma=1$, showing the results in Fig.~\ref{fig:ChaoticSpectrum}. For this case, we see the same overall behavior but with the eigenstate wave functions more spread over the Liouville basis, as can be seen in Figs.~\ref{fig:ChaoticSpectrum}~(d1) and~\ref{fig:ChaoticSpectrum}~(d2). Nevertheless, Fig.~\ref{fig:ChaoticSpectrum}~(d1) shows the projection of the eigenstate wave function more spread over the Liouville basis than the projection of Fig.~\ref{fig:ChaoticSpectrum}~(d2). This shows that the eigenstate wave function in chaotic regions has a very complex structure in Liouville space, and these projections are useful tools to understand it.

\section{SPECTRAL ANALYSIS AND QUANTUM CHAOS IN OPEN QUANTUM SYSTEMS}
\label{sec:SpectralAnalysis}

The way to perform the spectral analysis for open quantum systems with complex spectra was first outlined for periodically kicked dissipative tops in Ref.~\cite{Grobe1988}. These studies were latter extended to other open systems with finite dimension~\cite{Akemann2019,Hamazaki2020,Rubio2022}, which suggest that the behavior regarding regularity and chaos in open quantum systems is universal~\cite{Grobe1988,Akemann2019}. Thus, we follow the procedure exposed in these references to corroborate the validity of the GHS conjecture for the infinite-dimensional open Dicke model.

\subsection{Eigenvalue Spacing Distributions for Regular and Chaotic Complex Spectra}
\label{sec:ComplexSpacingDistribution}

In isolated quantum systems, the real eigenvalues of their Hamiltonians, $\epsilon_{k}\in\mathbb{R}$, can be ordered in increasing order, $\epsilon_{k}\leq\epsilon_{k+1}$. The spacing is defined as the separation between an eigenvalue $\epsilon_{k}$ and its nearest neighbor $\epsilon_{k+1}$, $s_{k}=\epsilon_{k+1}-\epsilon_{k}$. Performing an unfolding procedure of the spectrum~\cite{Guhr1998}, we can study its spectral fluctuations using the nearest-neighbor spacing distribution, which follow typically the Poisson distribution, $P_{\text{P}}(s)=\text{exp}(-s)$, for integrable (regular) systems and the Wigner-Dyson surmise, $P_{\text{WD}}(s) = (\pi/2)s\,\text{exp}(-\pi s^{2}/4)$, for the nonintegrable (chaotic) ones~\cite{Bohigas1984,Guhr1998}.

For open quantum systems, the Liouvillians are non-Hermitian and the eigenvalues are complex, $\varphi_{k}\in\mathbb{C}$, such that, the standard treatment to analyze spectral fluctuations is not longer applicable. For these systems, the spacing is understood as the minimal Euclidean distance in the complex plane for an eigenvalue $\varphi_{k}$ and its nearest neighbor $\varphi_{k}^{1\text{N}}$, $s_{k}=|\varphi_{k}-\varphi_{k}^{1\text{N}}|$. After performing an unfolding procedure for complex spectra (see Appendix~\ref{app:UnfoldingComplexSpectra} for an explanation of this technique), we can study, analogously to the isolated systems, the spectral fluctuations for open quantum systems.

Typically, the nearest-neighbor spacing distribution for integrable (regular) open quantum systems follows a 2D Poisson distribution~\cite{HaakeBook,Grobe1988,Akemann2019}, which is given by
\begin{equation}
    \label{eqn:2DPDistribution}
    P_{\text{2DP}}(s) = \frac{\pi}{2}s\,e^{-\pi s^{2}/4}.
\end{equation}
Note that this distribution is functionally the same as the Wigner-Dyson surmise, which characterizes the chaotic cases in isolated quantum systems.

However, for nonintegrable (chaotic) open quantum systems, the nearest-neighbor spacing distribution follows the distribution of the Ginibre unitary ensemble (GinUE)~\cite{HaakeBook,Ginibre1965,Grobe1988,Akemann2019,Hamazaki2020}, given by
\begin{equation}
    P_{\text{GinUE}}(s) = \prod_{k=1}^{\infty}\frac{\Gamma(1+k,s^{2})}{k!}\times\sum_{k'=1}^{\infty}\frac{2s^{2k'+1}e^{-s^{2}}}{\Gamma(1+k',s^{2})},
\end{equation}
where $\Gamma(k,z) = \int_{z}^{\infty}dt\,t^{k-1}e^{-t}$ is the incomplete Gamma function, $\int_{0}^{\infty} ds\,P_{\text{GinUE}}(s)=1$, and $\bar{s} = \int_{0}^{\infty} ds\,s\,P_{\text{GinUE}}(s)\approx1.1429$. To compare this distribution with numerical values, a scaling must be made to ensure that its first moment is unity
\begin{equation}
    \label{eqn:GinUEDistribution}
    \widetilde{P}_{\text{GinUE}}(s) = \bar{s}P_{\text{GinUE}}(\bar{s}s),
\end{equation}
with $\int_{0}^{\infty} ds\,\widetilde{P}_{\text{GinUE}}(s) = 1$ and $\int_{0}^{\infty} ds\,s\,\widetilde{P}_{\text{GinUE}}(s) = 1$.

In the limit $s\to0$, both distributions tend to the power law
\begin{equation}
    P_{\beta}(s) \propto s^{\beta},
\end{equation}
where the power $\beta=1,3$ identifies the degree of level repulsion, linear (regular) for integrable cases and cubic for nonintegrable (chaotic) ones, which seems to be universal in open quantum systems~\cite{HaakeBook,Grobe1988,Grobe1989}.

To corroborate that a data set comes from a given distribution, the well-known Anderson-Darling test can be implemented for the spacings $s_{k}=|\varphi_{k}-\varphi_{k}^{1\text{N}}|$, by computing the parameter~\cite{Anderson1952}
\begin{align}
    \label{eqn:ADTest}
    A^{2} = - N - \sum_{k=1}^{N}\frac{2k-1}{N}(\ln[F_{\text{X}}(s_{k})] + \ln[1-F_{\text{X}}(s_{N+1-k})]),
\end{align}
where the spacings are arranged in increasing order $s_{k}\leq s_{k+1}$, and $F_{\text{X}}(s) = \int_{0}^{s}ds'P_{\text{X}}(s')$ is the cumulative distribution function of the probability distribution $P_{\text{X}}(s)$ with X=2DP,GinUE. When the Anderson-Darling parameter is greater than a threshold, $A^{2}>2.5$, we can conclude with $95\%$ of confidence that the data set does not come from the given probability distribution.

\subsection{Ratio of Consecutive Eigenvalue Spacings for Complex Spectra}
\label{sec:ComlexSpacingRatio}

The ratio of consecutive eigenvalue spacings was introduced to study spectral fluctuations in isolated systems with real eigenvalues~\cite{Oganesyan2007,Atas2013}. The advantage of this measure is that the spectra can be studied without implementing unfolding procedures, which can be ambiguous in some cases. The last measure can be extended to open systems with complex eigenvalues. The procedure is detailed in Ref.~\cite{Sa2020}, where the complex ratio takes the form
\begin{equation}
    \label{eqn:ComplexRatio}
    Z_{k}=r_{k}e^{i\theta_{k}}=\frac{\varphi_{k}^{1\text{N}}-\varphi_{k}}{\varphi_{k}^{2\text{N}}-\varphi_{k}},
\end{equation}
where $\varphi_{k}^{1\text{N}}$ and $\varphi_{k}^{2\text{N}}$ are the first- and second-nearest neighbor of an eigenvalue $\varphi_{k}$, respectively.

The generic results from isolated quantum systems, where the eigenvalues of integrable quantum systems are uncorrelated (Poisson distribution) and those of nonintegrable ones show level repulsion (Wigner-Dyson surmise), have an analogy in open quantum systems. In open quantum systems, the sets of eigenvalues of integrable systems (2D Poisson distribution) are uncorrelated in the complex plane showing a flat (delocalized) distribution. However, the sets of eigenvalues of nonintegrable systems (GinUE distribution) shows cubic level repulsion, which manifests itself with a suppression of the distribution at the origin and small angles~\cite{Sa2020}.

The expectation values of $r_{k}=|Z_{k}|$ and $\cos(\theta_{k})=\text{Re}(Z_{k})/r_{k}$ can be computed using the marginal distributions from each distribution 2DP and GinUE. The following results are obtained $\langle r_{k}\rangle_{\text{X}} = 2/3, 0.74$ and $-\langle \cos(\theta_{k})\rangle_{\text{X}} = 0, 0.24$ with X=2DP,GinUE; which can be used as a benchmark to validate numerical results.

\begin{figure*}[ht]
\centering
\includegraphics[width=\textwidth]{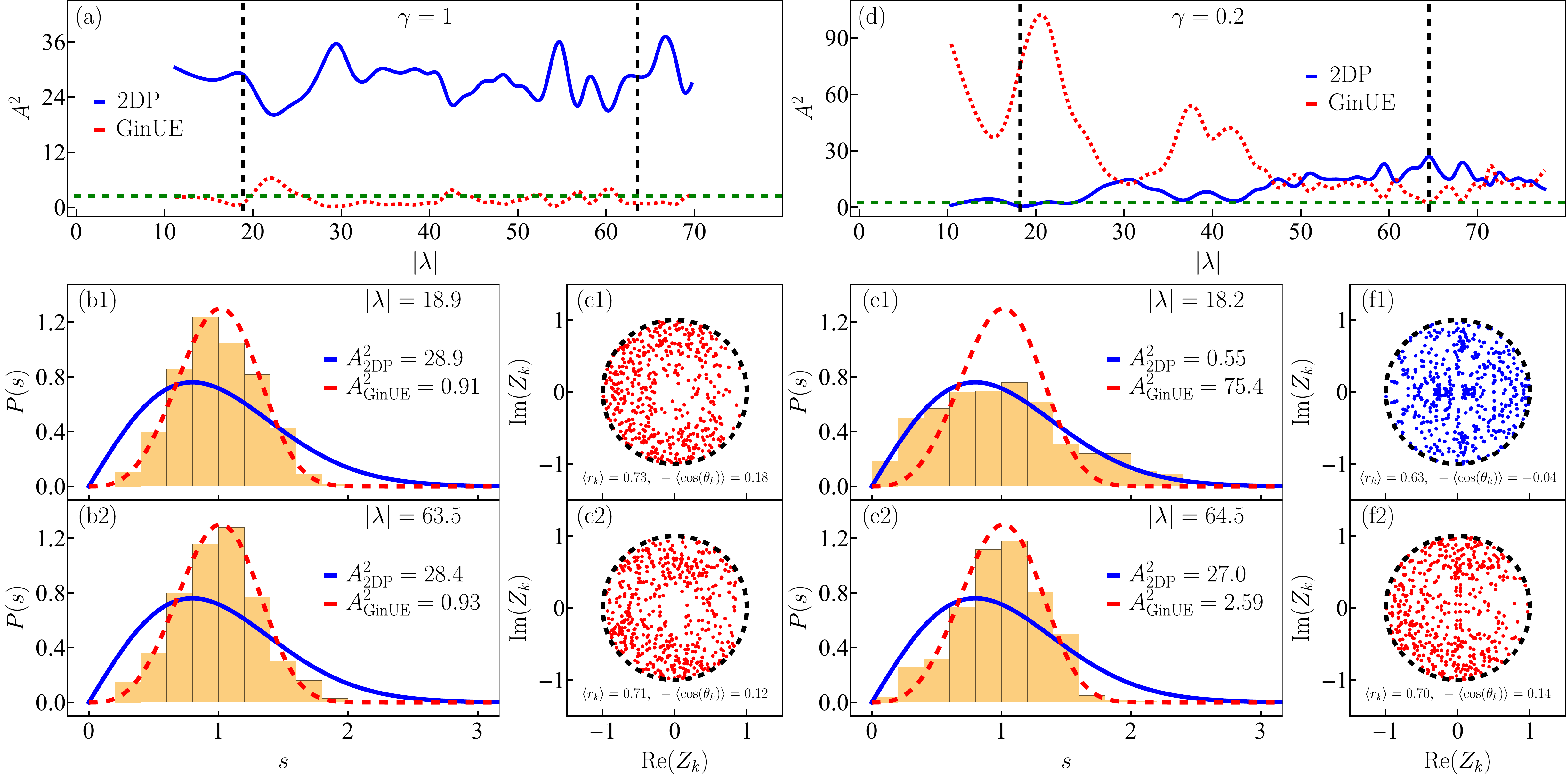}
\caption{(a) Anderson-Darling test for the complex spectrum with positive parity of the Dicke Liouvillian and the coupling strength $\gamma=1$ ($N_{\text{CES}}=9165$), which was computed numerically for the truncation value $n_{\max}=60$. The test was computed taking moving windows of the eigenvalue absolute value $|\lambda_{k}|$ (500 consecutive eigenvalues). The blue solid (red dotted) curve represents the Anderson-Darling parameter computed for the 2D Poisson (GinUE) distribution [see Eq.~\eqref{eqn:ADTest}]. The black dashed vertical lines represent the mean value of the eigenvalue absolute value of selected windows. The green dashed horizontal line represents the Anderson-Darling threshold, $A^{2}=2.5$. (b1, b2) Spacing distribution (bars) for the eigenvalues contained in the windows selected in panel (a), identified with black dashed vertical lines. The blue solid (red dashed) curve represents the 2D Poisson (GinUE) distribution [see Eqs.~\eqref{eqn:2DPDistribution}~and~\eqref{eqn:GinUEDistribution}]. (c1, c2) Complex ratio of consecutive eigenvalue spacings [red dots, see Eq.~\eqref{eqn:ComplexRatio}] for the eigenvalues contained in the same windows selected in panel (a). (d, e1, e2, f1, f2) The same as their corresponding panels (a, b1, b2, c1, c2), for the coupling strength $\gamma=0.2$ ($N_{\text{CES}}=11030$). The Liouvillian parameters in all panels are $\omega=\omega_{0}=j=\kappa=1$ and the tolerance value was chosen as $\Delta=10^{-3}$.}
\label{fig:ADTest_j1}
\end{figure*}

\begin{figure*}[ht]
\centering
\includegraphics[width=\textwidth]{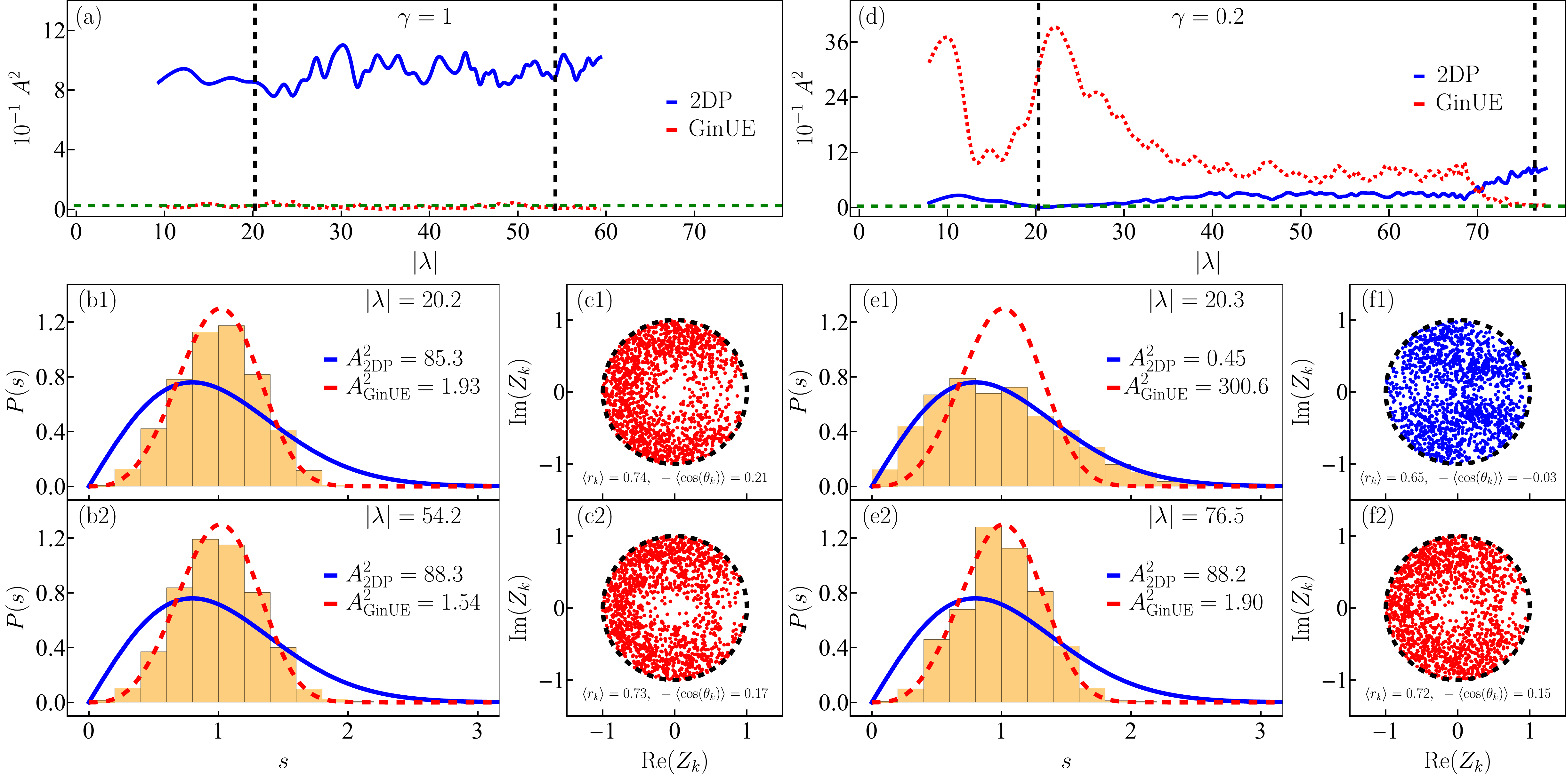}
\caption{The same as Fig.~\ref{fig:ADTest_j1} for the system size $j=3$ ($\mathcal{N}=6$ atoms). For this system size and the truncation value $n_{\max}=60$ were obtained $N_{\text{CES}}=36989,59001$ converged eigenstates (eigenvalues) for the coupling strengths $\gamma=1,0.2$ and the moving windows contain 1500 consecutive eigenvalues.}
\label{fig:ADTest_j3}
\end{figure*}

\section{CHAOS AND REGULARITY IN THE OPEN DICKE MODEL}
\label{sec:ChaosRegularityOpenDickeModel}

In this section we show numerical results characterizing the complex spectrum of the Dicke Liouvillian as chaotic or regular. We choose the spectrum computed with $n_{\max}=60$, which is the highest truncation value we achieved. For this case we get $N_{\text{CES}}=11030,9165$ converged eigenstates (eigenvalues) for the coupling strengths $\gamma=0.2,1$, with a tolerance value $\Delta=10^{-3}$ [see Figs.~\ref{fig:RegularSpectrum}~(a)~and~\ref{fig:ChaoticSpectrum}~(a), respectively].

First we present the case $\gamma=1$, whose isolated classical system shows chaotic behavior~\cite{Chavez2016}. To analyze this case, we have to take into account that, in absence of dissipation and in the superradiant phase ($\gamma>\gamma_{\text{c}}$), the real spectrum of the Dicke Hamiltonian transits from regularity (Poisson distribution) at low energies to chaos (Wigner-Dyson surmise) at high energies~\cite{Bastarrachea2014b,Bastarrachea2015}. This can be understood from the classical Dicke Hamiltonian, where at low energies the system can be approximated as a harmonic oscillator, while at high energies the system develops chaotic motion~\cite{Chavez2016}. As the open system could inherit some properties of the isolated system, we study the complex spectrum of the Dicke Liouvillian by regions.

To study the complex spectrum of the Dicke Liouvillian we take regions or windows of 500 consecutive eigenvalues, organized by the increasing absolute value of its eigenvalues [see Fig.~\ref{fig:ConvergedEigenvalues}~(b)], and apply the Anderson-Darling test [see Eq.~\eqref{eqn:ADTest}] to the spectrum, to check if the distribution comes from the 2D Poisson or the GinUE distribution [see Eqs.~\eqref{eqn:2DPDistribution}~and~\eqref{eqn:GinUEDistribution}]. The result is shown in Fig.~\ref{fig:ADTest_j1}~(a), where the blue solid curve is the Anderson-Darling test for the 2D Poisson distribution and the red dotted curve for the GinUE distribution, respectively. From this figure, it is clear that the complex spectrum of the Dicke Liouvillian is well determined by the GinUE distribution, confirming the chaotic behavior of the spectrum for high coupling strengths.

The last affirmation is corroborated by plotting the spacing distribution of the eigenvalues contained in two selected windows with mean value $|\lambda|=18.9,63.5$ in
Figs.~\ref{fig:ADTest_j1}~(b1) and~\ref{fig:ADTest_j1}~(b2), respectively. In both panels we can see that the eigenvalues contained in each window follow the GinUE distribution [see Eq.~\eqref{eqn:GinUEDistribution}], since the Anderson-Darling parameter does not cross the threshold $A^{2}=2.5$. Furthermore, we compute the complex ratio of consecutive eigenvalue spacings for the eigenvalues contained in the same windows [see Eq.~\eqref{eqn:ComplexRatio}], and plot them in Figs.~\ref{fig:ADTest_j1}~(c1) and~\ref{fig:ADTest_j1}~(c2). We can see in both panels that the point distribution is avoided at the origin as expected, and it looks fuzzy for small angles. The same panels show the numerical expectation values $\langle r_{k}\rangle$ and $-\langle\cos(\theta_{k})\rangle$ for each set of eigenvalues, which seem to agree with the theoretical expectation values from the GinUE distribution. The deviations are attributed to the low quantity of eigenvalues contained in each window, which must be suppressed when the system size increases; or instead, when the windows are wider containing more eigenvalues.

For the second case $\gamma=0.2$, the isolated classical system shows regular motion~\cite{Chavez2016}. For the real spectrum of the Dicke Hamiltonian with coupling strengths in the normal phase ($\gamma<\gamma_{\text{c}}$), the spectral fluctuations are generally regular (Poisson distribution)~\cite{Bastarrachea2014b}. Nevertheless,
we follow the same method of studying the complex spectrum of the Dicke Liouvillian taking regions organized by the increasing eigenvalue absolute value [see Fig.~\ref{fig:ConvergedEigenvalues}~(a)].

We take the same moving windows of 500 consecutive eigenvalues and apply the same procedure described above for the case $\gamma=1$. In Fig.~\ref{fig:ADTest_j1}~(d) we show the Anderson-Darling test. For low eigenvalue absolute values the complex spectrum follows the 2D Poisson distribution, confirming the regular behavior of the
spectrum. However, there is a transition region where this integrability breaks around $|\lambda|\sim25$. After this value, there are some fluctuations until the chaotic behavior of the spectrum seems to be reached at values around $|\lambda|\approx 65$. This suggests that the low coupling strength in the system does not guarantee the regularity of the system for the full complex spectrum.

As in the previous case, we plot the spacing distribution of the eigenvalues contained in two selected windows with mean value $|\lambda|=18.2,64.5$ in Figs.~\ref{fig:ADTest_j1}~(e1) and~\ref{fig:ADTest_j1}~(e2), respectively. Here, we can see that the eigenvalues contained in the first window follow the 2D Poisson distribution, while the second one seems to follow the GinUE distribution [see Eqs.~\eqref{eqn:2DPDistribution}~and~\eqref{eqn:GinUEDistribution}]. For the first window, the Anderson-Darling parameter does not cross the threshold $A^{2}=2.5$, while for the second window it is in the limit. Moreover, we plot in Figs.~\ref{fig:ADTest_j1}~(f1) and~\ref{fig:ADTest_j1}~(f2) the complex ratio of consecutive eigenvalue spacings for the eigenvalues contained in the corresponding windows. We can see in Fig.~\ref{fig:ADTest_j1}~(f1) that the point distribution is delocalized over the complex plane as expected, while in Fig.~\ref{fig:ADTest_j1}~(f2) the point distribution seems to be avoided at small angles, but not at all at the origin. We compute for both cases, the numerical expectation values $\langle r_{k}\rangle$ and $-\langle\cos(\theta_{k})\rangle$ for each set of eigenvalues. The first case seems to agree with the theoretical expectation values from the 2D Poisson distribution, while the second one with the GinUE distribution. The same deviations are attributed to the low quantity
of eigenvalues contained in each window.

Now, we make the previous analysis increasing the system size to corroborate our statements. We take the system size $j=3$ ($\mathcal{N}=6$ atoms) with the same truncation value $n_{\max}=60$, obtaining $N_{\text{CES}}=59001,36989$ converged eigenstates (eigenvalues) for the coupling strengths $\gamma=0.2,1$, with a tolerance value $\Delta=10^{-3}$.

In Fig.~\ref{fig:ADTest_j3} we show the results, first for the coupling strength $\gamma=1$ and then for $\gamma=0.2$. To perform the Anderson-Darling test, we take moving windows of 1500 consecutive eigenvalues. For the case $\gamma=1$, we can see in Fig.~\ref{fig:ADTest_j3}~(a) that the chaotic behavior of the complex spectrum of the Dicke Liouvillian is again confirmed, where the deviations of the Anderson-Darling parameter computed for the GinUE distribution decrease, showing an almost constant curve. Furthermore, in Figs.~\ref{fig:ADTest_j3}~(b1) and~\ref{fig:ADTest_j3}~(b2) we plot the spacing distribution of the eigenvalues contained in two selected windows with mean value $|\lambda|=20.2,54.2$, respectively. We see that not only the spacing distributions follow the GinUE distribution, but also the ratio of consecutive eigenvalue spacings, plotted in Figs.~\ref{fig:ADTest_j3}~(c1) and~\ref{fig:ADTest_j3}~(c2) for both windows, shows clearer the avoided regions at the origin and at small angles as expected. In the same way, we see that the agreement of the numerical expectation values $\langle r_{k}\rangle$ and $-\langle\cos(\theta_{k})\rangle$ with the theoretical ones improves as we have argued.

Taking the same moving windows of 1500 consecutive eigenvalues for the case $\gamma=0.2$, we can see in Fig.~\ref{fig:ADTest_j3}~(d) the Anderson-Darling test, which confirms the regular behavior of the complex spectrum of the Dicke Liouvillian at low eigenvalue absolute values. Furthermore, the transition to chaos is confirmed at high eigenvalue absolute values. This is an interesting feature of the open system, since the transition to chaos is developed slowly until the system behaves chaotic.

For the eigenvalues contained in two selected windows with mean value $|\lambda|=20.3,76.5$, we plot in Figs.~\ref{fig:ADTest_j3}~(e1) and~\ref{fig:ADTest_j3}~(e2) the spacing distribution. The first window follows the 2D Poisson distribution, while the second one follows the GinUE distribution, confirmed by the Anderson-Darling parameter. Moreover, the ratio of consecutive eigenvalue spacings for both windows is plotted in Figs.~\ref{fig:ADTest_j3}~(f1) and~\ref{fig:ADTest_j3}~(f2), showing the complete delocalization of the point distribution in the complex plane for the first window and the avoided regions at the origin and at small angles for the second one. As a matter of fact, we also see a better agreement of the numerical expectation values $\langle r_{k}\rangle$ and $-\langle\cos(\theta_{k})\rangle$ with the theoretical ones.

\section{SUMMARY AND CONCLUSIONS}
\label{sec:Conclusions}

We implemented a convergence criterion for the eigenstates and eigenvalues of the Dicke Liouvillian based on the eigenstate wave functions spread over the Liouville basis. The onset of chaos in the open Dicke model was successfully characterized by applying the standard spectral analysis proposed for open quantum systems.

For the high coupling strength case ($\gamma=1$), we detected the GinUE distribution for the eigenvalue spacings, typical for chaotic open quantum systems for all range of the eigenvalue absolute value of the complex spectrum. However, for the low coupling strength case ($\gamma=0.2$), we identified a richer structure, since at low eigenvalue absolute values we detected the 2D Poisson distribution for the eigenvalue spacings, typical for regular open quantum systems. Nevertheless, there is a regime where this integrability is broken and the onset of chaos arises in the system, implying that low coupling strengths do not guarantee the regularity of the system for all the spectrum.

We verified that the GHS conjecture is valid for the open Dicke model, confirming its universality for this infinite-dimensional system, when the spectral analysis of the Dicke Liouvillian is done by regions of its eigenvalues. We think that these studies are a first step to characterize the phenomenon of chaos in the open Dicke model. The analysis shown in this work can be extended adding other kinds of dissipative channels, as collective atomic decay or temperature effects. We think also, that the methods developed in this work could be extended to other open quantum systems with infinite Liouville space.

\section*{ACKNOWLEDGMENTS}

We thank Jorge G. Hirsch for the reading of this work and his valuable comments and suggestions. We acknowledge the support of the Computation Center - IIMAS, in particular to Adri\'an Chavesti. In the same way we acknowledge the support of the Computation Center - ICN, in particular to Enrique Palacios, Luciano D\'iaz, and Eduardo Murrieta. This work was supported by DGAPA-PAPIIT-UNAM under Grant No. IG101421 from Mexico. D.V. acknowledges financial support from the postdoctoral fellowship program DGAPA-UNAM.

\appendix

\section{DIAGONALIZATION OF THE DICKE LIOUVILLIAN}
\label{app:DiagonalizationDickeLiouvillian}

\subsection{Matrix Representation of the Dicke Liouvillian}

The way to diagonalize a Liouvillian is using the tetradic notation~\cite{MukamelBook}, where a matrix representation of the system can be obtained using an arbitrary basis of the form
\begin{equation}
    |k,l\rangle\rangle = |k\rangle\langle l|,
\end{equation}
where $|\bullet\rangle\rangle$ denotes a vector in the Liouville space composed by all the projectors of the Hilbert-space states $|\bullet\rangle$. For an $N$-dimensional basis of the Hilbert space, there will be an $N^2$-dimensional basis of the Liouville space.

Using the last procedure, the matrix representation of the Dicke Liouvillian takes the form
\begin{align}
    L_{k'l',kl}^{\text{D}} & = \langle\langle k',l'|\hat{\mathcal{L}}_{\text{D}}|k,l\rangle\rangle = \text{Tr}\{|l'\rangle\langle k'|\hat{\mathcal{L}}_{\text{D}}|k\rangle\langle l|\} \\
    & = \sum_{i}\langle i|l'\rangle\langle k'|\hat{\mathcal{L}}_{\text{D}}|k\rangle\langle l|i\rangle = L_{k'l',kl}^{\text{D},\gamma} + L_{k'l',kl}^{\text{D},\kappa}, \nonumber
\end{align}
where
\begin{align}
    L_{k'l',kl}^{\text{D},\gamma} & = -i\langle\langle k',l'|[\hat{H}_{\text{D}},\hat{\rho}]|k,l\rangle\rangle \\
    & = -i(\langle k'|\hat{H}_{\text{D}}|k\rangle\delta_{l,l'}-\langle l|\hat{H}_{\text{D}}|l'\rangle\delta_{k',k}), \nonumber
\end{align}
and
\begin{align}
    L_{k'l',kl}^{\text{D},\kappa} = & \kappa\langle\langle k',l'|(2\hat{a}\hat{\rho}\hat{a}^{\dagger}-\{\hat{a}^{\dagger}\hat{a},\hat{\rho}\})|k,l\rangle\rangle \\
    = & 2\kappa\langle k'|\hat{a}|k\rangle\langle l|\hat{a}^{\dagger}|l'\rangle + \nonumber \\
    & - \kappa(\langle k'|\hat{a}^{\dagger}\hat{a}|k\rangle\delta_{l,l'} + \langle l|\hat{a}^{\dagger}\hat{a}|l'\rangle\delta_{k',k}). \nonumber
\end{align}

\subsection{Fock Basis and Liouville Basis}

The standard way to diagonalize the Dicke Hamiltonian is using the Fock basis, which is composed by Dicke states $|j,m_{z}\rangle$ (with $m_{z}=-j,-j+1,\ldots,j-1,j$) for the atomic subspace and Fock states $|n\rangle$ (with $n=0,1,\ldots,\infty$) for the bosonic subspace in tensor product
\begin{equation}
    \label{eqn:FockBasis}
    |f\rangle = |n;j,m_{z}\rangle = |n\rangle\otimes|j,m_{z}\rangle,
\end{equation}
where the index $f(n,m_{z})=(2j+1)n+m_{z}+j+1$ reorders the elements of the basis with one value. As was mentioned previously, the Fock basis is infinite; nevertheless, a truncation finite value $n_{\max}$ of the bosonic subspace (maximum number of photons) is selected to solve the system numerically.

The Fock basis can be used to generate the diagonalization basis of the Dicke Liouvillian or Liouville basis
\begin{equation}
    \label{eqn:LiouvilleBasis}
    |f',f\rangle\rangle = |f'\rangle\langle f| = |n';j,m'_{z}\rangle\langle n;j,m_{z}|,
\end{equation}
with $m'_{z},m_{z}=-j,-j+1,\ldots,j-1,j$ and $n',n=0,1,\ldots,\infty$. Thus, the matrix elements of the Dicke Liouvillian are given by
\begin{align}
    \langle f'|\hat{H}_{\text{D}}|f\rangle = & (\omega n+\omega_{0}m_{z})\delta_{n',n}\delta_{m'_{z},m_{z}}  \\
    & +\frac{\gamma}{\sqrt{\mathcal{N}}}(\sqrt{n+1}\delta_{n',n+1}+\sqrt{n}\delta_{n',n-1}) \nonumber \\
    & \times(C_{m_{z}}^{+}\delta_{m'_{z},m_{z}+1} + C_{m_{z}}^{-}\delta_{m'_{z},m_{z}-1}), \nonumber
\end{align}
with $C_{m_{z}}^{\pm}=\sqrt{j(j+1)-m_{z}(m_{z}\pm1)}$, and
\begin{align}
    \langle f'|\hat{a}|f\rangle & = \sqrt{n}\delta_{n',n-1}\delta_{m'_{z},m_{z}}, \\
    \langle f'|\hat{a}^{\dagger}|f\rangle & = \sqrt{n+1}\delta_{n',n+1}\delta_{m'_{z},m_{z}}, \\
    \langle f'|\hat{a}^{\dagger}\hat{a}|f\rangle & = n\delta_{n',n}\delta_{m'_{z},m_{z}}, \\
    \langle f'|\hat{a}\hat{a}^{\dagger}|f\rangle & = (n+1)\delta_{n',n}\delta_{m'_{z},m_{z}}. 
\end{align}

\subsection{Liouville Basis with Well-Defined Parity}

The Fock basis $|f\rangle$ is an eigenbasis of the parity operator, $\hat{\Pi} = \text{exp}[i\pi(\hat{a}^{\dagger}\hat{a}+\hat{J}_{z}+j\hat{\mathbb{I}})]$,
\begin{equation}
    \hat{\Pi}|f\rangle = p|f\rangle,
\end{equation}
with eigenvalues $p = (-1)^{(n+m_{z}+j)} = \pm 1$, and allows us to select a basis with well-defined parity in the Hilbert space. The last feature allows us in the same way to select a basis with well-defined parity in the Liouville space, when the parity superoperator $\hat{\mathcal{P}}$ acts over the Liouville basis $|f',f\rangle\rangle=|f'\rangle\langle f|$
\begin{equation}
    \hat{\mathcal{P}}|f',f\rangle\rangle = \hat{\Pi}|f'\rangle\langle f|\hat{\Pi}^{\dagger} = P|f',f\rangle\rangle,
\end{equation}
with eigenvalues $P = (-1)^{(n'+m'_{z}-n-m_{z})} = \pm 1$.

\section{CONVERGENCE CRITERION OF EIGENSTATES OF THE DICKE LIOUVILLIAN}
\label{app:ConvergenceCriterionEigenstates}

\subsection{Eigenstates of the Dicke Hamiltonian}

The eigenstates of the Dicke Hamiltonian can be expanded in an arbitrary basis. Typically the Fock basis [see Eq.~\eqref{eqn:FockBasis}] is used to diagonalize the Dicke Hamiltonian, and the eigenstates of the system, $\hat{H}_{\text{D}}|E_{k}\rangle=E_{k}|E_{k}\rangle$, have the following representation:
\begin{equation}
    |E_{k}\rangle = \sum_{f=1}^{\mathcal{D}_{\text{H}}}c_{f}^{k}|f\rangle = \sum_{n=0}^{n_{\max}}\sum_{m_{z}=-j}^{j}c_{n,m_{z}}^{k}|n;j,m_{z}\rangle,
\end{equation}
where $n_{\max}$ is a truncation value of the bosonic subspace or maximum number of photons, $c_{f}^{k}=\langle f|E_{k}\rangle$ or $c_{n,m_{z}}^{k}=\langle n;j,m_{z}|E_{k}\rangle$ are the eigenstate wave function components, and the eigenstates are arranged in increasing order of their real eigenvalues $E_{k}\leq E_{k+1}$, $E_{k}\in\mathbb{R}$.

The probability to have $n$ photons in the eigenstate $|E_{k}\rangle$ is given by
\begin{align}
    \label{eqn:PhotonProbability}
    p_{n}^{k} = \sum_{m_{z}=-j}^{j}|\langle n;j,m_{z}|E_{k}\rangle|^{2} = \sum_{m_{z} = -j}^{j}|c_{n,m_{z}}^{k}|^{2},
\end{align}
and when this probability is evaluated for all the values $n=0,1,\ldots,n_{\max}$, it can be interpreted as a projection of the eigenstate wave function over the Fock basis. In this way, an eigenstate is considered well converged, when the projection of its wave function expanded over the Fock basis has zero probability for the maximum number of photons $n_{\max}$ (truncation value)~\cite{Bastarrachea2014PSb}
\begin{equation}
    p_{n_{\max}}^{k} = \sum_{m_{z}=-j}^{j}|c_{n_{\max},m_{z}}^{k}|^{2} \leq \delta,
\end{equation}
where $\delta$ is a tolerance value. The last statement can be interpreted alternatively as the eigenstate wave function is contained in the truncated Hilbert space, that is, all coefficients contributing to the wave function are contained inside the truncated Hilbert space. See a more detailed explanation of this eigenstate convergence method in Ref.~\cite{Bastarrachea2014PSb}.

\subsection{Eigenstates of the Dicke Liouvillian}

The convergence criterion explained above, which is valid for eigenstates of the Dicke Hamiltonian, can be extended for the eigenstates of the Dicke Liouvillian. By considering the Liouville basis [see Eq.~\eqref{eqn:LiouvilleBasis}] as the projectors of the Fock basis, we can diagonalize the Dicke Liouvillian, and the eigenstates of the open system, $\hat{\mathcal{L}}_{\text{D}}|\lambda_{k}\rangle\rangle=\lambda_{k}|\lambda_{k}\rangle\rangle$, take the form
\begin{align}
    \label{eqn:LiouvilleEigenstateExpansion}
    |\lambda_{k}\rangle\rangle & = \sum_{f',f=1}^{\mathcal{D}_{\text{H}}}c_{f',f}^{k}|f',f\rangle\rangle = \sum_{f'=1}^{\mathcal{D}_{\text{H}}}\sum_{f=1}^{\mathcal{D}_{\text{H}}}c_{f',f}^{k}|f'\rangle\langle f| \\
    & = \sum_{n',n=0}^{n_{\max}}\sum_{m'_{z},m_{z}=-j}^{j}c_{n',m'_{z},n,m_{z}}^{k}|n';j,m'_{z}\rangle\langle n;j,m_{z}|, \nonumber
\end{align}
where $n_{\max}$ is the same truncation value of the isolated bosonic subspace, $c_{f',f}^{k}=\langle\langle f',f|\lambda_{k}\rangle\rangle$ are the eigenstate wave function components, and the eigenstates are arranged in increasing order of their complex-eigenvalue absolute values $|\lambda_{k}|\leq|\lambda_{k+1}|$, $\lambda_{k}\in\mathbb{C}$.

Analogously to the isolated system, we can define an extension of Eq.~\eqref{eqn:PhotonProbability} for the eigenstate $|\lambda_{k}\rangle\rangle$ of the open system, such that, we have two weight distributions
\begin{align}
    \label{eqn:EigenstateWaveFunction1}
    P_{1,n'}^{k} = \sum_{n=0}^{n_{\max}}\sum_{m'_{z},m_{z} = -j}^{j}|c_{n',m'_{z},n,m_{z}}^{k}|^{2}, \\
    \label{eqn:EigenstateWaveFunction2}
    P_{2,n}^{k} = \sum_{n'=0}^{n_{\max}}\sum_{m'_{z},m_{z} = -j}^{j}|c_{n',m'_{z},n,m_{z}}^{k}|^{2},
\end{align}
which can be interpreted as projections of the eigenstate wave function over the Liouville basis for all the values $n',n=0,1,\ldots,n_{\max}$. For this case, an eigenstate is considered well converged, when both projections of its wave function expanded over the Liouville basis have zero contribution for the truncation value $n_{\max}$. Evaluating Eqs.~\eqref{eqn:EigenstateWaveFunction1} and~\eqref{eqn:EigenstateWaveFunction2} in $n',n=n_{\max}$, we get the expressions~\eqref{eqn:ConvergenceCriterion1} and~\eqref{eqn:ConvergenceCriterion2} shown in the main text.

\begin{figure}[ht]
\centering
\includegraphics[width=\columnwidth]{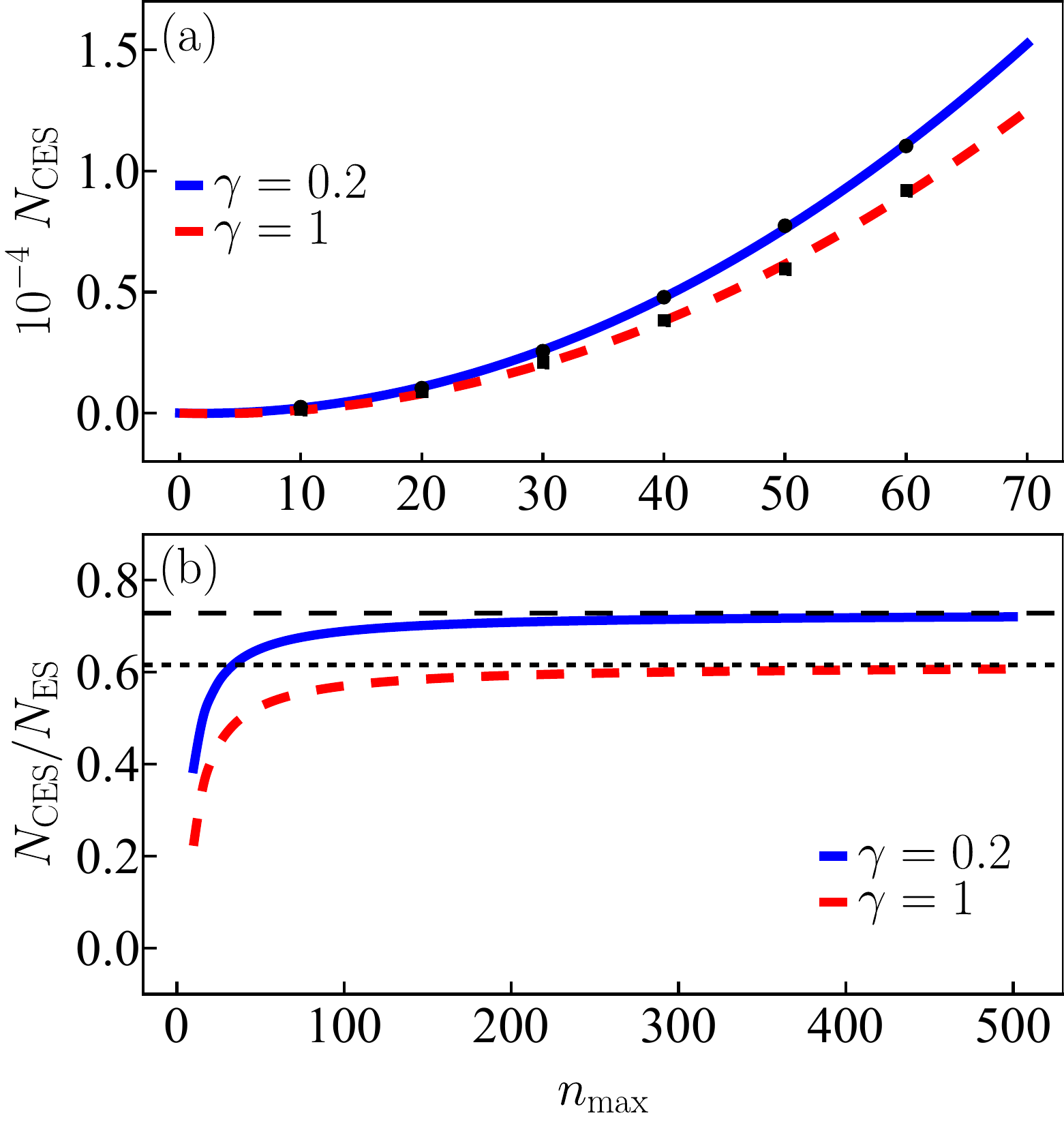}
\caption{Panel (a): Number of converged eigenstates $N_{\text{CES}}$ with positive parity of the Dicke Liouvillian for two coupling strength values $\gamma=0.2$ (black dots) and $\gamma=1$ (black squares), which were computed numerically for the truncation values $n_{\max}=10,20,30,40,50,60$. The blue solid (red dashed) curve depicts an analytical fit of the black dots (squares). Panel (b): Ratio of the number of converged eigenstates $N_{\text{CES}}$ to the total number of eigenstates $N_{\text{ES}}$ for the same coupling strength values $\gamma=0.2$ (blue solid curve) and $\gamma=1$ (red dashed curve). The black dashed (dotted) horizontal line represents the asymptotic value of the last ratios in the limit $n_{\max}\to\infty$ [see Eq.~\eqref{eqn:AsymptoticValue}]. The Liouvillian parameters in both panels are $\omega=\omega_{0}=j=\kappa=1$ and the tolerance value was chosen as $\Delta=10^{-3}$.}
\label{fig:ConvergedEigenstates}
\end{figure}

\subsection{Robustness of the Eigenstate Convergence Criterion}

For the well-defined parity Liouville basis, the dimension of the Liouville space $\mathcal{D}_{\text{L},P}$ (which defines the number of eigenstates $N_{\text{ES}}$) is given by
\begin{align}
    N_{\text{ES}} & = \mathcal{D}_{\text{L},P=\pm1} \\
    & = \left\{\begin{array}{ll}
         \mathcal{D}_{\text{H}}^{2}/2 & \text{if } (-1)^{n_{\max}}=-1, \\
         (\mathcal{D}_{\text{H}}^{2}\pm1)/2 & \text{if } (-1)^{n_{\max}}=1, 
    \end{array}\right. \nonumber
\end{align}
where $\mathcal{D}_{\text{H}}=(2j+1)(n_{\max}+1)$ is the dimension of the Hilbert space of the isolated system.

In Fig.~\ref{fig:ConvergedEigenstates}~(a) we show the number of well-converged eigenstates $N_{\text{CES}}$ selected under the eigenstate convergence criterion [see Eqs.~\eqref{eqn:ConvergenceCriterion1}~and~\eqref{eqn:ConvergenceCriterion2}] with a tolerance value $\Delta=10^{-3}$, for all the same truncation values $n_{\max}=10,20,30,40,50,60$ shown in the main text. By fitting the numerical results, we find a quadratic behavior for the number of converged eigenstates
\begin{equation}
    N_{\text{CES}} = A_{1}n_{\max}+A_{2}n_{\max}^{2},
\end{equation}
where $A_{1}=-11.49,-15.60$ and $A_{2}=3.28,2.77$ identify the fitting values for each coupling strength $\gamma=0.2,1$.

Using the last analytical expressions, we can find their asymptotic value in the limit $n_{\max}\to\infty$ for the ratio of converged eigenstates to the total number of eigenstates obtained in each implementation
\begin{equation}
    \label{eqn:AsymptoticValue}
    \lim_{n_{\max}\to\infty}\frac{N_{\text{CES}}}{N_{\text{ES}}} = \frac{2A_{2}}{(2j+1)^{2}}.
\end{equation}
For the parameters $A_{2}=3.28,2.77$ we find the asymptotic values $0.729$ and $0.616$, respectively. In Fig.~\ref{fig:ConvergedEigenstates}~(b) we show the ratio $N_{\text{CES}}/N_{\text{ES}}$ as a function of the truncation value $n_{\max}$, with their corresponding asymptotic value for the cases $\gamma=0.2,1$. We see in this figure that the fraction of converged eigenstates is bounded for both cases, tending asymptotically to a constant value in the limit $n_{\max}\to\infty$.

\section{UNFOLDING OF COMPLEX SPECTRA}
\label{app:UnfoldingComplexSpectra}

The unfolding of complex spectra is needed to remove system specific structures from it, in the same way as occurs in the real spectra, and can be implemented in different ways~\cite{HaakeBook,Markum1999,Akemann2019,Hamazaki2020}. Following the method presented in Ref.~\cite{Akemann2019}, the spectral density of states can be separated in an average (system specific) and a fluctuating (universal) part
\begin{equation}
    \nu(\varphi_{k}) = \sum_{l=1}^{N}\delta^{(2)}(\varphi_{k}-\varphi_{k,l}) = \nu_{\text{a}}(\varphi_{k})+\nu_{\text{f}}(\varphi_{k}),
\end{equation}
where the averaged spectral density of states is approximated by a sum of Gaussian functions near each complex eigenvalue $\varphi_{k}\in\mathbb{C}$ of a set with $N$ elements
\begin{equation}
    \nu_{\text{a}}(\varphi_{k}) \approx \frac{1}{2\pi\sigma^{2}N}\sum_{l=1}^{N}e^{-|\varphi_{k}-\varphi_{k,l}|^{2}/(2\sigma^{2})},
\end{equation}
where $\sigma = 4.5S$, $S = N^{-1}\sum_{k=1}^{N}s_{k}$, and the spacings $s_{k} = |\varphi_{k}-\varphi_{k}^{1\text{N}}|$ are scaled as
\begin{equation}
    \widetilde{s}_{k} = \frac{\sqrt{\nu_{\text{a}}(\varphi_{k})}}{\widetilde{S}}s_{k},   
\end{equation}
with $\widetilde{S} = N^{-1}\sum_{k=1}^{N}\sqrt{\nu_{\text{a}}(\varphi_{k})}s_{k}$.

\bibliography{main}

\end{document}